\documentclass[final]{raa}            

\usepackage{graphicx,times}             
\usepackage{natbib}
\usepackage{amssymb,amsmath}
\usepackage{gensymb}
\usepackage{epstopdf}
\usepackage{physics}
\usepackage{multirow}
\usepackage{ulem}
\usepackage{soul} 
\bibpunct{(}{)}{;}{a}{}{,}

\usepackage[pagebackref=true]{hyperref}
\hypersetup{
    colorlinks=true,
    linkcolor=blue,
    filecolor=magenta,
    urlcolor=blue,
    citecolor=blue,
    pdfpagemode=FullScreen,
    }

\newcommand{\cm}{\ensuremath{\,{\rm cm}}}

\newcommand{\pc}{\ensuremath{\,{\rm pc}}}

\newcommand{\ms}{\ensuremath{\,{\rm ms}}}

\newcommand{\Jy}{\ensuremath{\,{\rm Jy}}}
\newcommand{\mJy}{\ensuremath{\,{\rm mJy}}}

\renewcommand{\Re}{\ensuremath{\,{\rm Re}}}

\begin{document}

\title{The FRB-searching pipeline of the Tianlai Cylinder Pathfinder Array}

   \volnopage{Vol.0 (20xx) No.0, 000--000}      
   \setcounter{page}{1}          

   \author{Zijie Yu 
      \inst{1,2}
   \and Furen Deng
      \inst{1,2}
         \and Shijie Sun
      \inst{1,3}
   \and Chenhui Niu
      \inst{4}
   \and Jixia Li
      \inst{1,3}
   \and Fengquan Wu
      \inst{1,3*}
    \and Wei-Yang Wang 
    \inst{2}
    \and Yougang Wang
    \inst{1,3}
    \and Shifan Zuo
    \inst{1,3}
    \and Lin Shu
    \inst{5,6}
    \and Jie Hao 
    \inst{5,6}
    \and Xiaohui Liu
    \inst{1,2}
    \and Reza Ansari
    \inst{7}
    \and Ue-Li Pen
    \inst{8,9}
    \and Albert Stebbins
    \inst{10}
    \and Peter Timbie
    \inst{11}
   \and Xuelei Chen
      \inst{1,2,3,12,13\dag}
   }

   \institute{National Astronomical Observatories, Chinese Academy of Sciences,
             Beijing 100101, China; *{\it wufq@bao.ac.cn}, \dag{\it xuelei@cosmology.bao.ac.cn}\\
        \and
             University of Chinese Academy of Sciences Beijing 100049, China;\\
        \and
            Key Laboratory of Radio Astronomy and Technology, Chinese Academy of Science, Beijing 100101, China;\\
        \and
            Institute of Astrophysics, Central China Normal University, Wuhan 430079, China;\\
        \and 
            Institute of Automation, Chinese Academy of Sciences, Beijing 100190, China;\\
        \and
            Guangdong Institute of Artificial Intelligence and Advanced Computing, Guangzhou 510535, China;\\
        \and
            Universit\'e Paris -- Saclay, CNRS/IN2P3, IJCLab, 91405 Orsay, France;\\
        \and
            Institute of Astronomy and Astrophysics, Academia Sinica, Roosevelt Road, Taipei 10617, Taiwan, China;\\ 
        \and
            Canadian Institute for Theoretical Astrophysics, University of Toronto, 60 Saint George Street, Toronto, ON M5S 3H8, Canada;\\
        \and
            Department of Physics, University of Wisconsin -- Madison, Madison, Wisconsin 53706, USA;\\
        \and
            Fermi National Accelerator Laboratory, P.O. Box 500, Batavia IL 60510--5011, USA;\\
        \and
            Department of Physics, College of Sciences, Northeastern University, Shenyang 110819, China;\\
        \and 
            Center of High Energy Physics, Peking University, Beijing 100871, China;\\
\vs\no
   {\small Received 20xx month day; accepted 20xx month day}}

\abstract{This paper presents the design, calibration, and survey strategy of the Fast Radio Burst (FRB) digital backend and its real-time data processing pipeline employed in the Tianlai Cylinder Pathfinder array. The array, consisting of three parallel cylindrical reflectors and equipped with 96 dual-polarization feeds, is a radio interferometer array designed for conducting drift scans of the northern celestial semi-sphere. The FRB digital backend enables the formation of 96 digital beams, effectively covering an area of approximately 40 square degrees with 3 dB beam. Our pipeline demonstrates the capability to make automatic search of FRBs, detecting at quasi-real-time and classify FRB candidates automatically. The current FRB searching pipeline has an overall recall rate of 88\%.
During the commissioning phase, we successfully detected signals emitted by four well-known pulsars: PSR B0329+54, B2021+51, B0823+26, and B2020+28. We report the first discovery of an FRB by our array, designated as FRB 20220414A. We also investigate the optimal arrangement for the digitally formed beams to achieve maximum detection rate by numerical simulation.
\keywords{instrumentation: miscellaneous, techniques: interferometers}
}

   \authorrunning{Zijie Yu et al. }            
   \titlerunning{The FRB-searching pipeline of the Tianlai Cylinder Pathfinder Array}  

   \maketitle


%
%
%
%
%
%
%

\section{Introduction}
FRBs has become a focus of current research in radio astronomy since its initial discovery of the Lorimer burst \citep{Lorimer2007,Thornton2013}. FRBs are believed to originate mostly from sources beyond our galaxy, as they exhibit a dispersion measure (DM) that exceeds what is expected from the interstellar medium within our own galaxy along the same line of sight. This hypothesis was confirmed when FRB 20121102A was localized to a host galaxy with a redshift of $z=0.193$ \citep{Bassa17,Chatterjee17,Marcote17}. 
Thousands of FRB bursts have been detected by radio telescopes worldwide, more than 790 FRB sources have been published, with a majority ($\sim 70\%$) by CHIME \citep{CHIME202112}, including 66 repeaters. 
Despite the unknown origin and radiation mechanism of FRBs, theses observations have significantly advanced our understanding of this enigma. The Blinkverse database \citep{Blinkverse2023} gathers the information for about 7900 radio bursts from these sources. For an overview of FRB properties and the progress in identifying their sources and corresponding astrophysical mechanisms see e.g. \citet{2021Univ....7..453C}.

FRB signals typically last for several to tens of milliseconds, necessitating telescopes equipped with sub-millisecond sampling backends for detection. Beamforming techniques are also commonly employed in FRB searches to enhance sky coverage and localization. As a consequence, a significant volume of data is generated. There are two approaches to tackle this issue. One option is to allocate an adequate number of storage servers. Alternatively, real-time data processing can be implemented, storing only the data containing potential FRB candidates while discarding the remaining data.

This paper presents the design and commissioning results of the digital FRB backend for the Tianlai Cylinder Pathfinder array. The Tianlai experiment consists of a dish pathfinder array \citep{Wu2021dish} and a cylinder pathfinder array \citep{Li:2020ast,Zhang_2016,Zuo:2018,Sun_2022,Li_2021}, which aim to test key technologies for cosmological surveys using the 21-cm intensity mapping method \citep{Chen2012,Xu_2014}. The cylinder pathfinder array consists of three reflectors, and is equipped with a total of 96 dual-polarization receivers. The basic parameters and configuration of the array are shown in Table \ref{tab:cylinder_properties} and Figure \ref{fig:layout}, respectively. Initially, these arrays were equipped with correlators that provided visibilities with a second-level sampling rate \citep{Niu2019}, which is sufficient for sky map reconstruction \citep{ZUO2021100439}. However, the cylinder array has a relatively wide field of view, covering approximately 96 square degrees in its --3 dB primary beam, which allows it to be used for blind search of FRB, once a high speed digital backend is equipped. In this paper we introduce the design, implementation and commissioning test of the digital transient backend of the cylinder array, which is added to run in parallel with the original digital correlator backend. Similarly, a digit backend for the dish array has also been built and installed, though it handles fewer inputs and form only 16 beams \citep{Yu:2022RAA}.

This paper is organized as follows. In Section 2, 
a brief review of the antenna and analog electronic system of the Tianlai cylinder pathfinder array, we describe the beam forming principle and its hardware implementation, followed by the transient searching pipeline. System operation and performance is described in Section 3, including calibration and test of digital beam forming efficiency, and a study of the sensitivity of the system. 
We have characterized the system and the detection pipeline by observing bright pulsars and through 
mock FRB injection. Beam arrangement optimization is discussed in Section 4 and we report the detection of one FRB event during our test observation campaign in Section 5. Finally, we conclude in Section 6. 

\begin{table}[b!]
    \centering
    \caption{Basic parameters of the Tianlai Cylinder Pathfinder Array}
    \begin{tabular}{c c}
    \hline
         Parameters & Values \\
    \hline 
         Number of cylinders  & 3            \\
         Reflector E-W diameter & 15 m              \\
         Reflector N-S length & 40 m           \\
         Number of receivers  & 96       \\
         f/D & 0.32                      \\
         Average SEFD & 67.18 kJy        \\
         Latitude & 44.15\degree N       \\
         Longitude & 91.80\degree E      \\
         Current Observing frequency & 685--810 MHz          \\
    \hline
    \end{tabular}
    \label{tab:cylinder_properties}
\end{table}

\section{System Design}

\subsection{Analog System of the Array}

\begin{figure}
    \centering
    \includegraphics[width=0.6\textwidth]{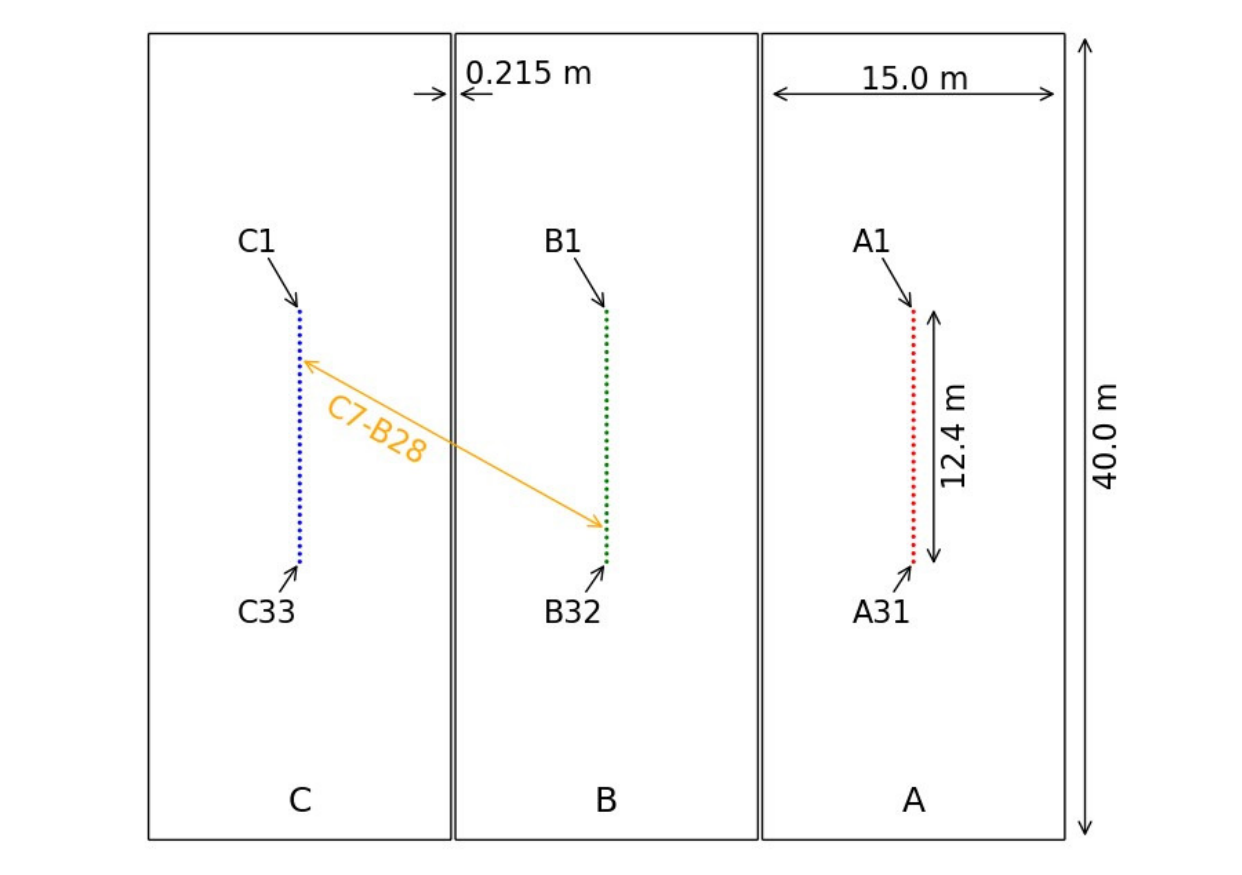}
    \caption{The configuration of Tianlai cylinder pathfinder reflectors and the ayout of receiver feed numbers \citep{Li:2020ast}. Each rectangle represents a cylinder reflector. The geometric parameters, including the length (40 m), width (15 m) of the reflector and the gap between them (0.215 m) are annotated. The dots in the center of each cylinder marks the position of a dual-polarization feed, which are denoted by the letter A, B, C with a sequential number. A baseline is labelled by its pair of feeds, e.g. the baseline made by feed C7 and B28 is labelled as C7-B28.}
    \label{fig:layout}
\end{figure}

The analog part of the cylinder array is unchanged, and a detailed description has been given in \citet{Li:2020ast}. In Table \ref{tab:20220414_properties} we list its main parameters. Three cylinder reflectors, designated as the A, B, and C cylinder, are adjacent to each other, with axis in the north-south direction. Receiver feeds are arranged in a uniformly spaced linear array along the focus line of each cylinder, designated A1, A2, ... C33, with the same maximum distance between the two ends of the array (12.4 meters), but with slightly different number of feeds on each cylinder (31, 32, 33 respectively), to suppress grating lobes of the array \citep{Zhang_2016}. An interferometric baseline can be specified by the designation of the pair of feeds (Figure \ref{fig:layout}). The 96 dual linear polarization receiver feeds generates a total of 192 outputs of analog radio signals. These are then amplified by low noise amplifiers (LNAs) located on the feeds. The amplified signals are converted to optical signals and transported via optical fiber cable to a station where the digital facilities are housed. At the digital facilities, the optical signals are converted back to radio signals and undergo bandpass filtering to select a 100 MHz bandwidth, which is currently set to 700--800 MHz. The signals are then down-converted to an intermediate frequency (IF) range of 135--235 MHz using a mixer. Subsequently, each of the IF signal is split into two outputs, then amplified to compensate for the reduced power, and fed to the original digital correlator and the new digital transient backend. Both the original digital correlator and the new FRB backend actually digitize the IF signal within 125--250 MHz independently, which are mapped to its original frequency in the data processing.

\subsection{Beamforming Principle}

\begin{figure*}
    \centering
    \includegraphics[width=0.8\textwidth]{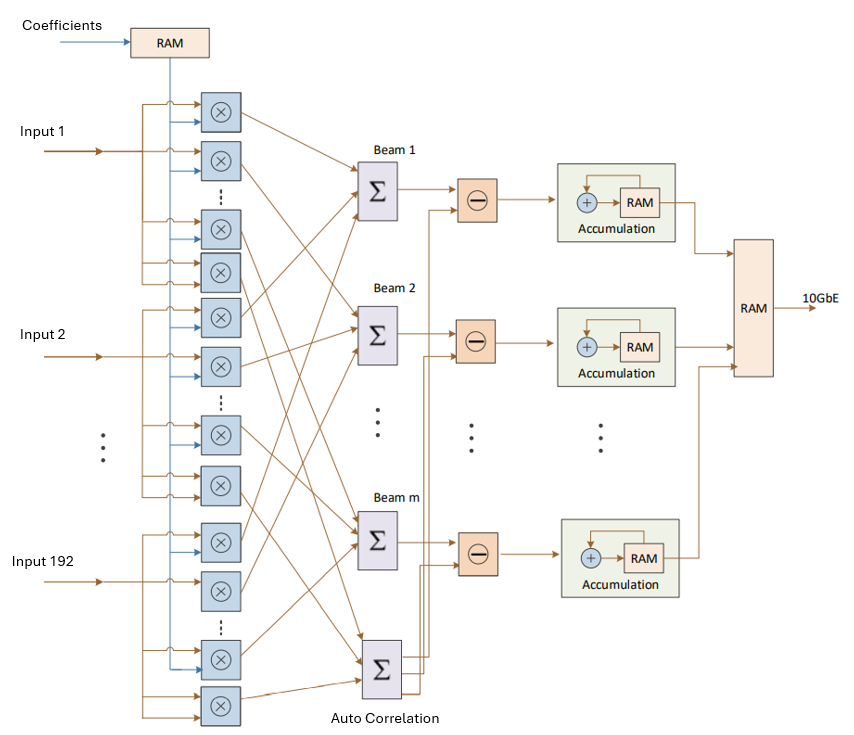}
    \caption{The principle of beamforming. Input signal is firstly digitized and then multiplied with complex coefficients, then added up to form beams. The auto-correlation of all inputs are summed and can be subtracted from beams to reduce noise. The beam data is accumulated according to the required time resolution.}
    \label{fig:beamforming}
\end{figure*}

Figure \ref{fig:beamforming} 
illustrates the principle of beamforming. 
To form a digital beam, we can add up the voltage output from each array element with a complex weight, 
\begin{equation}
J (\bm{k}) = \sum_a w_a(\bm{k})  \mathcal{E}_a
\label{eq:sum}
\end{equation}
where $w_a(\bm{k})$ denotes the complex weight of antenna $a$ for beam direction $\bm{k}$.
The voltage of element $a$ is given by  
\begin{equation}
\mathcal{E}_a = g_a \int e^{-2\pi j {\bf n}\cdot {\bf u}_a} A_a({\bf n}) E({\bf n}) d^2{\bf n} + \eta_a, 
\label{eq:voltage}
\end{equation}
where $A_a({\bf n})$ describes the voltage response for antenna $a$, $\bm{u}_a \equiv \bm{x}_a/\lambda$ is the position vector of the antenna in unit of wavelength, the integration is over sky directions, and $g_a$ is the complex gain of the instrument for unit $a$, $\eta_a$ represents the noise for that channel. We write $g_a = |g_a| e^{j \phi_a}$ to extract the overall delay of the signal transport chain. The complex weights are set as 
\begin{equation}
w_a(\bm{k})= \frac{1}{|g_a|} e^{-j \phi_a} e^{2\pi j \bm{k} \cdot \bm{u}_a},
\label{eq:weight}
\end{equation}
then the phases of all antennas are equal for the direction $\bm{n}=\bm{k}$.  
The power of this beam is given by
\begin{eqnarray}
S (\bm{k}) &\propto & \langle |J (\bm{k})|^2 \rangle \nonumber \\ 
&=&\sum_{a,b} \int A^*_a(\bm{n}) A_b(\bm{n}) I(\bm{n}) e^{-2\pi j(\bm{n}_a-\bm{k})\cdot \bm{u}_{ab}}  d^2\bm{n} + \sum_{a} |g_a|^{-2} \langle |\eta_a|^2 \rangle
\end{eqnarray}
where $\bm{u}_{ab}=\bm{u}_b-\bm{u}_a$, and $I(\bm{n})$ is the sky radiation intensity in direction $\bm{n}$.
If a burst occurs in the direction $\bm{q}$ at some moment, and assume that sky in other directions and the receiver noise remains nearly constant at this time, then the received signal is given by $ S= S_0 + \Delta S $,
where 
\begin{equation}
\Delta S =\sum_{ab} e^{-2\pi j(\bm{q}-\bm{k})\cdot \bm{u}_{ab}} A^*_a(\bm{q}) A_b(\bm{q}) I (\bm{q}),    
\label{eq:DeltaS}
\end{equation} 
and $S_0$ is the S value without burst. Multiple beams can be formed at the same time, and for each of the beam, we can search for radio pulses with dispersion delay. This is realized by the beamformer system, which perform the digital beam-forming, and the beam data is then streamed to de-dispersion servers, which searches for the dispersed radio pulse in the data at real time. 

\subsection{The Beamformer}

The beamformer consists of an F engine which  performs Fast Fourier Transform (FFT) to convert the data from the time domain to the frequency domain, and a B engine which forms beams at each frequency by a weighted summation of the signals from different feeds. We have a total of 96 dual linear polarization feeds, generating a total of 192 signal inputs. Each of these are digitized by the F-engine and connected to the B-engine, but due to limited processing power, currently we are only forming 96 digital beams, and for this we choose to use a single polarization. The choice of signal input can be made at real time by specifying the weight, and we are now usually using the $x$-polarization, i.e. the one aligned in the North-South direction, which has less cross-coupling between the feeds. The weight of the unused channels are set to zero.

\begin{figure*}
    \centering
    \includegraphics[width=0.9\textwidth]{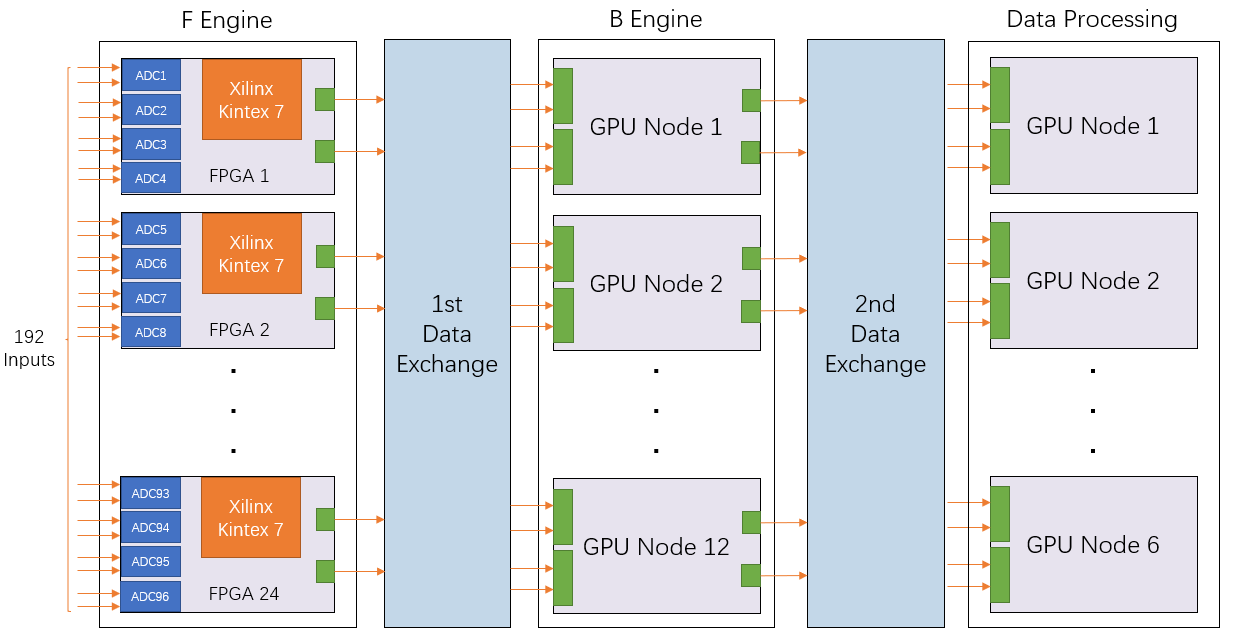}
    \caption{The hardware configuration of the beamformer. The F engine  consists 24 FPGA boards, each is equipped with 4 ADCs and a Xilinx Kintex-7 FPGA to digitize 8 inputs. Each green box in the figure represents a QSFP port for data transfer. The first data exchange is accomplished with two 48-port switchboard. The B engine consists 12 GPU nodes, each has two GPU servers. A 48-port switchboard is used to process the second data exchange. The 96 beams are searched using 6 data processing GPU servers.}
    \label{fig:beamformer}
\end{figure*}

\begin{figure*}
    \centering
    \includegraphics[width=0.8\textwidth]{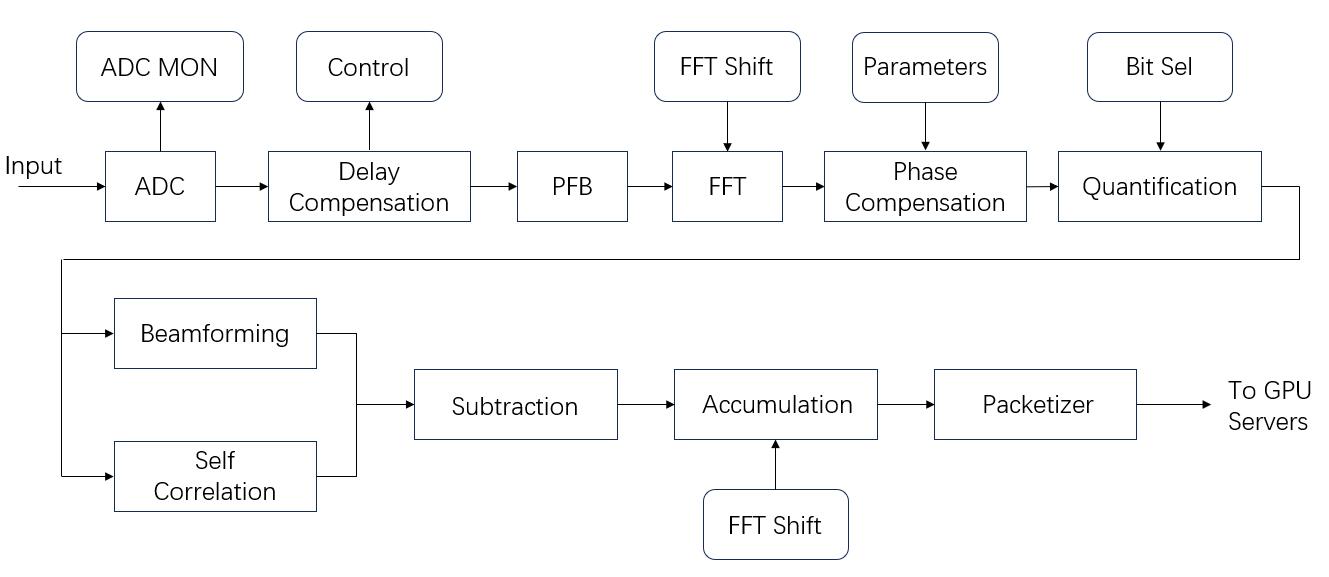}
    \caption{The workflow of the beamformer.}
    \label{fig:beamformer_workflow}
\end{figure*}

The hardware setup of the system is shown in Figure \ref{fig:beamformer}.
The F engine is implemented on 24 sets of analog-to-digital (ADC) converter boards, each digitize 8 analog signals and export data through two QSFP ports. The B engine is based on 12 GPU nodes, each handling  a block of frequency channels for all beams formed out of the digital signals. The data is transferred from the F engine to the B engine through the 1st Data Exchange which includes two 48-port switchboards, each handling half of the frequencies. The 1st Data Exchange re-package the Fourier transformed data from different receivers according to their frequency, with the same frequency to the same processing unit in the B engine. The output of the B engine goes through the 2nd Data Exchange, which includes another 48-port switchboard. The 2nd Data Exchange packages the data of different frequencies according to their beam, thus generating 96 beams, which are transferred to the de-dispersion nodes for further data processing.

\begin{figure*}
    \centering
    \includegraphics[width=0.8\textwidth]{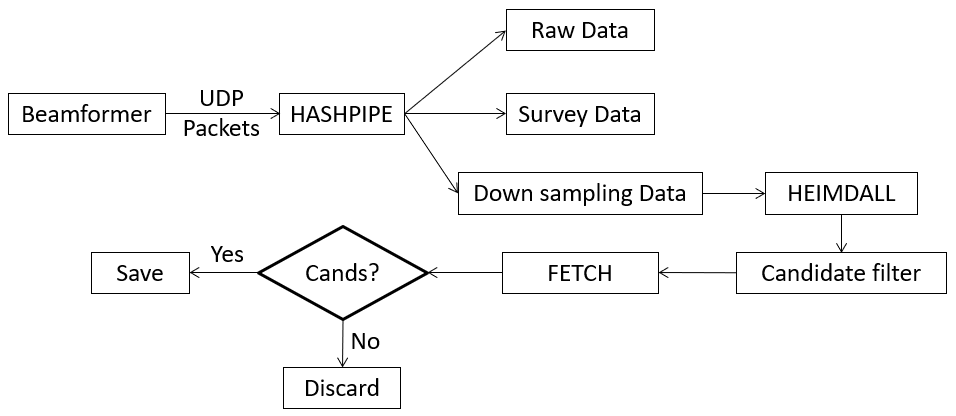}
    \caption{The workflow of the transient searching pipeline. The HASHPIPE is utilized to process the UDP packets from beamformer and generate three data streams. The raw data stream store full time and frequency resolution data. The survey data stream is accumulated to a 10.06 second integration interval for sky map reconstruction. The down sampling data reduces both time and frequency resolution to 1/4 and is for FRB quasi-real-time searching. We utilized the HEIMDALL program, candidate filter and the FETCH program to search and classify FRB candidates. }
    \label{fig:pipeline_workflow}
\end{figure*}

In Figure \ref{fig:beamformer_workflow} we show the work flow of the beamformer. The digitized data from the ADC is transformed to the frequency domain by FFT. Prior to the FFT, the data is first pre-processed. This includes a shifting of a few blocks of data in time for each signal channel separately to compensate for different time delays, which arise from the long optical fibers connecting the antenna to the analog and digital electronics in the station house. To mitigate leakage and minimize side lobes, a polyphase Filter Bank (PFB) is employed, which uses a Hamming function with 4 taps and 2048 samples. Following the PFB, we proceed with a 2048-sample FFT on the pre-processed data. After the FFT, a phase compensating for the residue instrument delay is added. The time block shift prior to the PFB and the phase shift after FFT are determined by calibration measurement and set by the system control, the compensating phase can be changed by command. Once these are done, the data are cut off to the required number of bits, then the beams are formed by summing the results with the appropriate complex weighting factors. An auto-correlation for each channel is also calculated, which can be subtracted subsequently to reduce the average output. Each of the beams is then integrated (accumulated) to yield the output at millisecond level, the results are packaged to sent to the GPU servers for dispersion and pulse search.

\subsubsection{F-engine}

Each ADC board is equipped with a Xilinx Kintex-7 FPGA, four TI ADS5407 ADCs, a 72 MB QDR, two QSFP ports for high-speed data transmission, and an Ethernet port for control and configuration. The ADCs on each board sample signals from 2 feeds at a sampling rate of 250 Msps.

To achieve synchronization for the ADC boards, a \texttt{sync} signal is sent to all boards. The rising edge of the \texttt{sync} signal triggers all the boards to simultaneously reset their status and clock.  
Shortly after this, the control starts transmitting a \texttt{trigger} signal, 
its rising edge triggers the boards commence writing data into a first-in-first-out (FIFO) at the next rising edge of the \texttt{syncout} signal. When the falling edge of the \texttt{trigger} signal is detected, the boards output the data from the FIFO. Thus, the output data is synchronized. 

\subsubsection{Switchboard}

To establish the 1st Data Exchange system with  96 ports, two switchboards are utilized, each is equipped with 48 SFP28 10 GbE ports and 8 QSFP28 100 GbE ports. These switchboards are interconnected through the use of 5 QSFP28 ports. Prior to transmission to the B engine, data from the F engines undergo rearrangement. Specifically, data from different input feeds with identical frequencies are grouped together, enabling the servers within the B engine to form beams at each frequency with all the data needed. 

Similarly, the data outputted by the B engine is also rearranged by the 2nd Data Exchange system, which is realized with one switchboard of 48 ports, so that the different frequencies of the same beam is directed to the same processing node for de-dispersionn and pulse search.

\subsubsection{B engine}

Each GPU node consists two servers and each server is equipped with an NVIDIA RTX 3060 and 8 GB VRAM. The workflow of the B engine is illustrated in Figure \ref{fig:beamforming}. Within this workflow, data from different frequency channels and feeds are weighted by a coefficient matrix, followed by summation to generate the 96 beams. Subsequently, the data is accumulated to achieve a time resolution of 0.1 ms. 
Lastly, the B engine send the data to the de-dispersing nodes.

\subsection{Transient Searching pipeline}

The beamformed stream data is processed on six GPU servers, each is equipped with two Intel Xeon E5-2699 V3 CPUs, two NVIDIA RTX 3080 GPUs, and 314 GB RAM. Each GPU server is connected to the beamformer via four 10GbE cables, with each cable carrying data for four beams. The workflow of the transient searching pipeline is shown in Figure \ref{fig:pipeline_workflow}. The data is transmitted from the beamformer in UDP format. At the GPU server end, the High Availability Shared Pipeline Engine (HASHPIPE)\footnote{https://casper.astro.berkeley.edu/wiki/HASHPIPE} is utilized to receive data packets and write them into SIGPROC\footnote{https://sigproc.sourceforge.net/} Filterbank format files, with each file containing 81.2-second data from one beam.

Once a data file is generated, we employ HEIMDALL\footnote{https://sourceforge.net/projects/heimdall-astro/}, a GPU-accelerated single pulse searching software \citep{Barsdell2012}, to perform de-dispersion and identify transient candidates within the file. The candidates events identified by HEIMDALL are then subjected to a candidate filter, which applies multiple criteria to eliminate obvious radio frequency interference (RFI). The remaining candidates are classified using FETCH\footnote{https://github.com/devanshkv/fetch}, a deep learning-based transient classifier \citep{Agarwal2020}. If any of the candidates are classified as potential FRBs, the pipeline saves data files from all beams to disk and sends notification emails to administrators. In the absence of any candidates, the pipeline discards the raw data and candidate searching files to free up disk space. The processing time may vary slightly but is generally shorter than the observing time covered by the file. 

Note that this part of the pipeline works independently from the beamforming pipeline, so in principle its hardware or software can also be replaced, as long as the alternative system can handle the data and complete the pulse-searching task in comparable or shorter time. It is conceivable that in the future, by employing faster algorithm, e.g. the Hough transform method \citep{Zuo:2020}, the processing cycle time could be further shortened.

\subsubsection{Data Stream}

Four instances of the HASHPIPE program are created on each server to process data from one cable. Each instance generates three child threads. The net child thread reads data from the packets and writes them into ring buffer 1. The \texttt{cal} thread reads data from ring buffer 1, reshapes the data array into SIGPROC's Filterbank format, and writes them into ring buffer 2. The \texttt{output} thread reads data from ring buffer 2 and writes them into Filterbank format data files.

The search for pulses in beamformed raw data, with a time and frequency resolution of 0.098304 ms and 0.122 MHz respectively, incurs significant computational load. To reduce this load, the output thread divides the raw data stream from the beamformers into three separate data streams, each serving a different purpose. The first data stream is dedicated to storing the raw data to preserve all information in case of a FRB detection. The second data stream undergoes downsampling by a factor of 4 in both time and frequency, resulting in data with a resolution of 0.393 ms and 0.488 MHz. All quasi-real-time processes mentioned below are performed on this downsampled data stream. The third data stream is accumulated to a time resolution of 10.06 seconds, which can be utilized for imaging with this backend.

\subsubsection{Calibrator Noise Source}
The Tianlai array utilizes a noise source to perform relative phase calibration. This calibrator noise source is activated periodically for a few seconds every few minutes, depending on the specific settings. However, the presence of these noise signals leads to the identification of numerous candidates when using HEIMDALL for single pulse searches. This significantly increases the workload of the pipeline and obscures genuine signals that may be present in the vicinity of the noise. To mitigate this interference, we apply a zero DM removal procedure to the data, by subtracting the frequency-averaged value at each sample point.

\subsubsection{De-dispersion and pulse searching}
\label{sec:de-dispersion and pulse searching}

Two parallel HEIMDALL threads are executed on each GPU server to process data from 16 beams. The HEIMDALL configuration is presented in Table \ref{tab:HEIMDALL}. Burst searches are conducted over a range of DM from 0 to 1500 $\rm pc\,cm^{-3}$, with a default SNR loss tolerance of 1.25. This results in a total of 377 DM trials. The output time interval is 0.393 milliseconds, while the maximum width of the boxcar filter (rectangular in time domain) is set to 128 samples, corresponding to a pulse width of 50.3 milliseconds.  According to the Blinkverse FRB catalogue \citep{Blinkverse2023}, the width of most FRBs fall within this range: among 3418 recorded FRB pulse widths (including FAST and CHIME samples), only 15 of them are smaller than 0.393 ms, and 7 of them are larger than 50.3 ms. 

 If a burst occurs near the end of the data block, the delayed low frequency part of the burst may not fall within the present data block, then it could be missed in the search of the present block.  We solve this end-of-file problem by copying the 8 seconds of data at the end of the file to the start of the next data file, so that such missed pulse can still be detected at the next block.
To mitigate the impact of RFIs, frequency channels that are heavily contaminated are masked using the 'zap channel' option of HEIMDALL. The masked frequency is set empirically based on past observations and is fixed in FRB-search during all observations, while data within these bands will still be saved for off-line studies.

\begin{table}
    \centering
    \caption{HEIMDALL parameters}
    \begin{tabular}{c c}
    \hline
         Parameters & Values \\
    \hline 
         DM range & 0--1500 $\rm pc\,cm^{-3}$       \\
         SNR loss tolerance & 1.25 \\
         DM trials & 377              \\
         Maximum boxcar width & 128 samples   \\
         Pulse width & 0.393--50.3 ms   \\
         Zap channels & 0--140 , 246--256 \\
         Masked Frequency & 741.6--810.0 , 685.0--689.9 MHz \\
    \hline
    \end{tabular}
    \label{tab:HEIMDALL}
\end{table}

After the search, HEIMDALL produces a candidate file containing the identified potential candidates, each accompanied by information such as the SNR (Signal to Noise Ratio), candidate time, DM, and boxcar width. Upon completion of processing for all beams, the HEIMDALL-provided {\it coincidencer} script is utilized to summarize the candidates. This script groups together candidates with matching times and similar DM values from different beams, effectively consolidating them into a single candidate. Additionally, the {\it coincidencer} script records the number of beams in which a candidate is detected and identifies the beam with the highest SNR for each candidate.

\subsubsection{Candidates filtering}
HEIMDALL produces a substantial number of candidates, ranging from hundreds to thousands per minute. However, the majority of these candidates are attributed to background noise fluctuations, the noise source used for calibration, and RFI.  We implement a filtering process to reject the non-astronomical candidates based on several criteria. 

Firstly, we discarded faint candidates below a certain threshold, at present this threshold is set at SNR = 8. Additionally, candidates with a boxcar width of 0 or exceeding 7 are also discarded, as they are likely to be RFIs with pulse widths either too small ($<0.393$ ms) or too large ($>50.3$ ms) compared to typical FRBs at 750 MHz.

Secondly, candidates detected in more than 20 beams are excluded, as astronomical signals are expected to be detected only in beams pointing towards nearby directions. Strictly speaking, this criterion should depend on how the beams are placed, as the digitally formed beams can be directed at any direction, and they could in principle overlap with each other, where the number of detected beam may not be a good criterion. However, currently we have placed the beams apart from each other in our observations, and this criterion rejects mostly nearby RFI which affecting all beams.  

In cases where a pulse is detected in more than one beams, HEIMDALL reports multiple candidates, one for each beam, which unnecessarily increases the computational cost during classification. Thus, when such simultaneous candidates are present, the one with the highest SNR are used for candidate classification described below. Note that if a pulse is classified as real, the data corresponding to that time frame is saved, and in the later off-line processing the detection in multiple beams could be utilized in analysis.

\begin{figure*}
    \centering
    \includegraphics[width=0.8\textwidth]{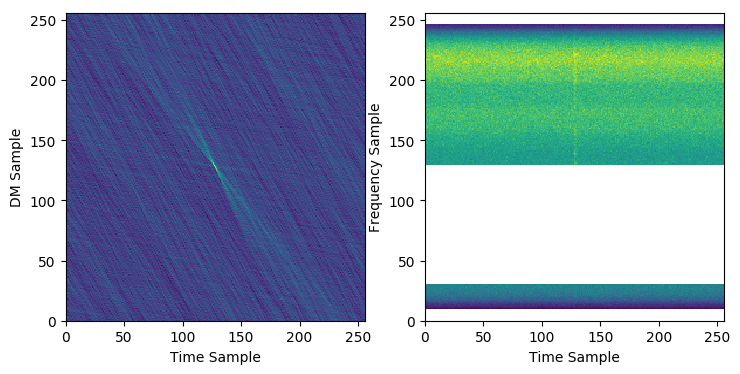}
    \caption{An example of FETCH input using the data of FRB 20220414A. The left panel shows the DM-time matrix, the right panel shows the frequency-time matrix, for the time frame which contains the burst. The figure shows a total of 256 DM values and 256 time values, centered at the detected event. The 256 frequency is down-sampled from our original 1024 frequency channels. The white region is contaminated by RFIs and masked.}
    \label{fig:fetch_candidate}
\end{figure*}

\begin{table*}
    \centering
    \caption{\texttt{FETCH} Models and their performance tested with real ASKAP and Parkes FRBs and RFIs \citep{Agarwal2020}. The number in brackets after each model is the number of unfrozen layers. $k$ is the fusion hyperparameter. The Models a, c, and f are used in this work.}
    \resizebox{\textwidth}{!}
    {
    \begin{tabular}{c c c c c c c}
    \hline
         \multirow{2}{*}{Label} & \multirow{2}{*}{FT Model} & \multirow{2}{*}{DMT Model} & \multirow{2}{*}{$k$} & ASKAP FRBs & Mislabelled ASKAP & Mislabelled Parkes \\
         & & & & (/33) & RFIs (/10639) & RFIs (/478)\\
    \hline 
         a & DenseNet121 (4)  & Xception (21) & 256 & 33 & 2 & 0\\
         b & DenseNet121 (4)  & VGG16 (2)     & 32  & 28 & 5 & 48\\
         c & DenseNet169 (11) & Xception (21) & 112 & 33 & 16 & 6\\
         d & DenseNet201 (7)  & Xception (21) & 32  & 33 & 12 & 29\\
         e & VGG19 (4)        & Xception (21) & 128 & 33 & 16 & 7\\ 
         f & DenseNet169 (11) & VGG16 (2)     & 512 & 33 & 2  & 1  \\
         g & VGG19 (4)        & VGG16 (2)     & 128 & 29 & 1  & 10 \\
         h & DenseNet201 (7)  & InceptionResNetV2 (34) & 160 &33 & 43 &40\\
         i & DenseNet201 (7)  & VGG16 (2)     & 32  & 33 &15 & 52\\
         j & VGG19 (4)        & InceptionResNetV2 (34) & 512 & 33 & 33 & 3\\
         k & DenseNet121 (4)  & InceptionV3 (31) & 64 & 33 & 7 &70 \\
    \hline
    \end{tabular}
    }
    \label{tab:fetch_model}
\end{table*}

\subsubsection{Candidate Classification}
We currently use an artificial intelligence program \texttt{FETCH} to 
further classify the remaining candidates after the filtering process. 
The \texttt{FETCH} program uses different combination of Convolutional Neural Network (CNN) models to process the Frequency-Time (FT) and DM-Time (DMT) matrices, and replacing the top classification layer of both models with a dense layer to generate ensemble models, borrowing techniques developed in the image processing community \citep{Agarwal2020}.  

The training data set of \texttt{FETCH} is chosen from three different telescope backends, consisting simulated FRBs, real pulsar bursts, and real RFIs. The validation data set includes 6664 real pulsar bursts and 7319 RFIs. Real FRBs and RFIs observed with ASKAP, Parkes and BL are used to evaluate the model performance.

\texttt{FETCH} provides a script (\texttt{candmaker.py}) which 
reads the candidate files generated by our pipeline, and generates two data matrices: one for DMT and another for FT, as illustrated in Figure \ref{fig:fetch_candidate}. The resulting data matrices are then stored in an HDF5 file. Subsequently, the FETCH \texttt{predict.py} script reads all the HDF5 files in this batch, then utilizes the selected models to assess the likelihood of each candidate being an astronomical signal, and output the classification results in a CSV file. 

We combine the \texttt{FETCH} model a, c and f to classify the candidates by summing their predictions. These three models have an 100\% recall rate when tested with the real ASKAP FRBs and have relatively small amount of mislabelled RFIs. Models b and g missed several ASKAP FRBs, while models d, h, i, j, k performed poorly against RFIs, so these are not used. From model e and f, we choose model f to introduce diversity in the DMT model.

Based on the predicted probability, the candidates are classified into three groups: RFI, pulsar, and potential FRB.  Candidates with an average likelihood below a threshold value of 0.4 are discarded as RFIs. This threshold is chosen to recover most of the pulses in PSR B0329+54 observation. Candidates with an average likelihood above 0.4 and a DM value exceeding 50, are labeled as potential FRBs. Those with DM less than 50 but SNR$>100$ are also classified as potential FRBs, as such pulses are not frequently produced by pulsars. Pulses with an average likelihood above 0.4, DM value less than 50, and SNR below 100 are classified as pulsars. In the event of identifying a potential FRB candidate, our pipeline promptly sends a notification email to operators for further human inspection.

\section{System operation and performance}

The digitally formed beams can be pointed at any direction by adjusting the phase of the complex weighting factor. Currently, we have arranged the beams in a pattern similar to our array itself: the 96 beams are put in 3 columns along the sky meridian circle, i.e. in north-south direction. The separation between columns and the separation within each column are adjustable parameters, which are optimized for maximum detection rate. The area near the zenith would generally have a higher detection rate as it corresponds to the peak of the antenna primary beam. 

\subsection{Calibration}
\label{sec:Calibration}

To obtain the complex gain $g_a= |g_a| e^{j \phi_a}$ needed for beam forming, we make a calibration observation of a bright radio source. We form beams which include only a pair of feeds, with one of these a reference feed, and the other the feed to be calibrated, while the weight for all other feeds are all set to zero, in this way we determine the phase difference of each feed with respect to the reference feed.  As the beamformer allows forming multiple beams at the same time, we can determine all the instrument phases in one observation. For feeds C1, ..., C33 and B17, ..., B32, we use A1 as the reference feed, while for feeds B1, ..., B16 and A1, ..., A31 we use C33 as reference feed, so that only the longer baselines, which is more sensitive to phase variation and has less cross-couplings, are used for calibration. The A1-C33 measurement determines the phase difference between these two reference points, and all other phases can then be derived with respect to one of them.  

\begin{figure}
    \centering
    \includegraphics[width=0.45\textwidth]{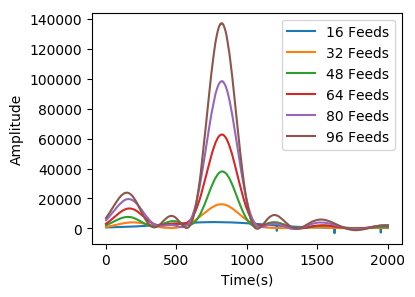}
    \caption{Synthesized beam test: beam amplitude vs. time during a Cas A transit, each line represents a beam ormed with different number of inputs as labelled. As we add more feeds, the amplitude of the peak increases and the beam becomes narrower.} 
    \label{fig:beamforming_amplitude}
\end{figure}

\begin{figure}
    \centering
    \includegraphics[width=0.45\textwidth]{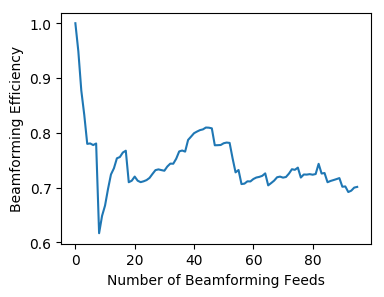}\\
    \caption{The efficiency of the synthesized beam, i.e. the ratio of the measured amplitude and the theoretical amplitude as a function of the number of feeds.}
    \label{fig:beamforming_eff}
\end{figure}

We model the primary beam as a Gaussian function, the output of the beam is fitted as 
\begin{eqnarray}
    \Re I_{ab}(t) = c_1^2\left[\cos(\Delta\varphi_{ab}(t) + \phi_{ab}) + c_2^2\right]\exp\left(\frac{\Delta\varphi_{ab}^2(t)}{2\sigma^2}\right) + D t,
    \label{eq:fit}
\end{eqnarray} 
where $\Delta\varphi_{ab}(t) = \varphi_{ab}(t) - \varphi_{ab}(t_{\rm transit})$, $\varphi_{ab} = 2\pi (\bm{n}-\bm{k})\cdot \bm{u}_{ab}$ and $t_{\rm transit}$ is the time for the source transit, $c_1, c_2$ are some constants. The $D\Delta t$ is a phenomenal term to capture the time variation of the noise bottom during the period of transit observation. The instrument phase difference $\phi_{ab}=\phi_a-\phi_b$ is then determined.

In ideal calibration scenarios, the intensity of the signal of a synthesized beam is $n^{2}$ times that of a single feed when the source is positioned at the beam's center. However, practical limitations, such as errors in the complex gain of the feeds, diminish the efficiency of beamforming as the number of feeds $n$ increases. To assess this efficiency, we conduct a test by observing Cas A after calibrating the complex gains using Cas A as a reference. In this test, the number of feeds used is equal to $n$, with beam $\#$1 formed using $n=1$, i.e. only one feed, the weight for other feeds are all set to zero; beam $\#$2 using $n=2$, and so forth. 

The transit of Cas A for these beams are shown in Figure \ref{fig:beamforming_amplitude}. For the first 32 beams we use the feeds from the same cylinder, while for more than 32 we also include feeds from different cylinders. We can see that with the increase of the number of feeds, the amplitude increases, showing that the beam-forming produce higher sensitivity than individual feed measurements. The beams narrowed down significantly when the feeds from the different cylinders are included.
There is also a peak prior to the main peak in this transit. It is at partly due to some asymmetric feature in the primary beam response, as we found similar peak in the transit of other sources such as Cyg A, though the variation of sky intensity may also contribute.

In Figure \ref{fig:beamforming_eff} we show the efficiency of the synthesized beam, i.e. the ratio of the measured peak amplitude and theoretical amplitude, as a function of the number of feeds used. If the amplitude and phase of each feed is exactly the same, or calibrated without error, the peak amplitude would be proportional to the number of feeds being used. However, as there are calibration errors, the actual amplitude is less than the theoretical one. In the figure we see once the additional feeds are added, the efficiency drops below 100\%, but with the available feeds added in, the efficiency stabilises at about 70\%.

\subsection{Sensitivity}
\label{sec:sensitivity}

The sensitivity of the system can be determined from a calibration observation of a known source, e.g. Cygnus A. After masking the RFI, data of 424 channels are used for calibration, ranging from 689.9 to 741.6 MHz, and the same band is used in FRB search. 
During its transit, we can see an increase in the received power,  induced by the transit of Cyg A, obtained by subtracting the peak value and the background value,
\begin{equation}
\Delta_{\rm cal}\equiv T_{\rm pk}-T_{\rm bg}= 236.92~\sigma_T,
\end{equation}
where $\sigma_T$ is estimated for the 689.9--741.6 MHz band (total bandwidth 51.7 MHz), a sampling time of 0.098304 ms, and a beam formed with the 96 feeds. The peak amplitude of the transit is obtained by averaging the data for 5 seconds, while the background is taken at about 20 minutes before the transit.

The flux of calibrator source (Cyg A) at this frequency is $S_{\rm cal}=2980 \Jy$, and $\Delta_{\rm cal} \propto  S_{\rm cal} B_{\rm cal}$, where $B_{\rm cal}$ is the primary beam gain in the direction of calibrator, so the sensitivity at a direction $\bm{n}$ for a single integration time is given by 
\begin{equation}
S_{n0} = S_{\rm cal} \frac{B_{\rm cal}}{B_n} \frac{\sigma_T}{\Delta_{\rm cal}}= 12.58 \left(\frac{B_{\rm cal}}{B_n}\right)~ \Jy
\label{eq:Sn0}
\end{equation}
where $B_n=B(\bm{n})$ is the primary beam gain in that direction. 

\begin{figure}
    \centering
    \includegraphics[width=0.5\textwidth]{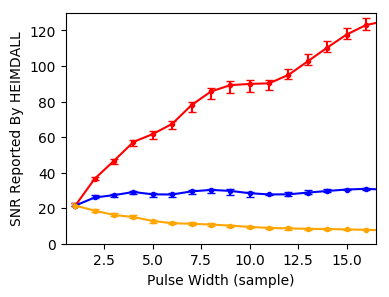}
    \caption{The SNR of mock FRBs as a function for pulse width, for a constant flux density of 280 Jy (red), a constant fluence of 280 Jy ms (orange), and a constant $F/\sqrt{\tau \Delta t}$ of $280 \Jy \ms^{1/2}$ (blue). The error bar is derived from 5 trials.  
    }
    \label{fig:fluence_strength}
\end{figure}

In pulse searching, the detection limit is usually given in fluence, which depends both on the pulse flux and the duration, $F=\int S_\nu (t) dt$, which has a unit of $\Jy \cdot \ms$. Denoting the sampling time interval as $\tau$, duration of the pulse $\Delta t$,  If $\Delta t$ is less than a single integration time interval $\tau$, then $S_n(\Delta t)=S_{n0}$. For longer interval, $S_n(\Delta t)= S_{n0} \sqrt{\tau/\Delta t}$, so 
\begin{equation}
    {\rm SNR} (\Delta t)=
    \left \{
    \begin{aligned}
     \frac{S}{S_{n0}} \frac{\Delta t}{\tau}=\frac{F}{S_{n0}~ \tau}, \quad \Delta t \textless \tau \\
     \frac{S}{S_{n0}} \sqrt{\frac{\Delta t}{\tau}} =\frac{F}{S_{n0} \sqrt{\tau \cdot \Delta t}},\quad \Delta t \geq \tau\\        
    \end{aligned}
    \right.
    \label{eq:SNR_dt}
\end{equation}
In the actual case, most bursts will have $\Delta t >\tau$, and we search through the maximum SNR using a boxcar filter for different width in time. Pulses with SNR greater than certain threshold are then selected. This is demonstrated in Figure \ref{fig:fluence_strength},
where we show the variation of SNR for different pulse width,  for the case with a constant flux (280 Jy), the case with a constant fluence ($280 \Jy \ms$), and the case with flux $\times \Delta t^{1/2}$ kept constant ($280 \Jy \ms^{1/2}$).  As predicted by Equation (\ref{eq:SNR_dt}), 
 the observed SNR of burst is proportional to $F/\sqrt{\tau \Delta t}$.

\subsection{Mock FRB test of Pipeline Performance}
\label{sec:Pipeline_Performance}

To examine our pipeline performance, we inject mock FRB signal into real-time data, and feed these to our transient detection pipeline. The pulse we generate are box-shaped, with flux ranges from 10 to 310 \Jy, pulse width between 0.8 to 24.8 ms, corresponding to 2 to 62 samples time, and DM between 70 to 1570 $\pc\cm^{-3}$. The combination of these parameters produces a total of 7488 mock FRBs.

We inject these mock FRBs into real search data during blind search by HASHPIPE. We inject 9 mock FRBs every 81.2 seconds, separated by approximately 10 seconds and into different beams. It takes 18.7 hours to inject all 7488 mock FRBs, which allows us to study the influence of day and night. To obtain the performance of each step in the pipeline, we show statistics for the recovered FRBs at three stages: after the HEIMDALL’s single pulse searching, after filtering, and after the FETCH classification (final stage of the pipeline). We analyze the result using recall rate and precision, given by equation (\ref{eq:recall}) and (\ref{eq:precision}). Recall rate shows our pipeline’s ability to detect potential FRBs, while precision indicates its performance against RFIs.

\begin{equation}
  {\rm Recall\,Rate}= \frac{\rm True\,Positive}{\rm True\,Positive + False\,Nagative}
\label{eq:recall}
\end{equation} 

\begin{equation}
  {\rm Precision} = \frac{\rm True\,Positive}{\rm True\,Positive + False\,Positive}
\label{eq:precision}
\end{equation} 

\begin{figure}
    \centering
    \includegraphics[width=0.8\textwidth]{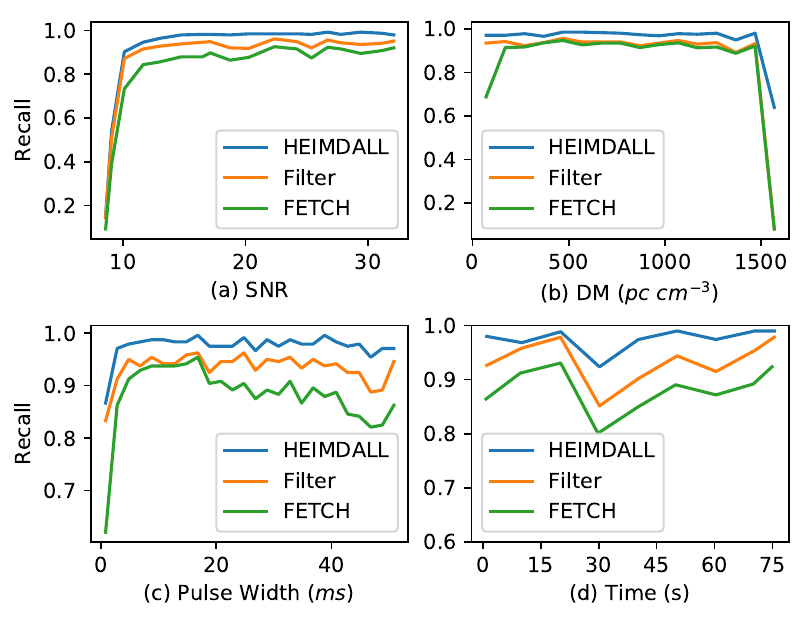}
    \caption{The recall rates of the mock FRBs recovered by our pipeline after each processing step as a function of SNR, DM, pulse width, and candidate time in each data file. For each plot the other parameters are kept at the average value. }
    \label{fig:recall}
\end{figure}

Figure \ref{fig:recall} shows how the recall rate depends on each parameters of the mock FRB, after marginalizing other parameters. The performance of our pipeline as measured by the recall rate depends on the Signal-to-Noise Ratio (SNR)  of the event. For FRBs with an SNR below 10, the recall rate is less than 0.5. However, for brighter pulses, the recall rate is above 0.8 and remained stable.

Our pipeline maintained a stable recall rate of approximately 0.9 across a wide range of DM values. Notably, the FETCH has a low performance at the two ends of DM values. In terms of pulse width, the recall rate drops at the narrow pulse end, at a pulse width of 0.8 ms. For longer burst duration, both HEIMDALL and the filter performed normally, while FETCH exhibited a 5$\%$ decrease in performance. Figure \ref{fig:recall}(d) displays the recall rate of simulated FRBs injected at different candidate times within each data file, which shows that our pipeline consistently delivered stable performance, and  we found no noticeable difference between day and night. Note here we have cut off the last 6 seconds of the data, where the detection efficiency drops due to the reason explained in Section \ref{sec:de-dispersion and pulse searching}, but this is not a problem of FETCH, and since the same segment of data will be searched at the beginning part of the next time block where such drop would not occur,  this does not matter.  
\begin{table}
    \centering
    \caption{Overall Recall and Precision of pipeline}
    \begin{tabular}{c c c c c c}
    \hline         
         \multirow{2}{*}{Step} & \multicolumn{2}{c}{All Mock FRBs} &\multicolumn{2}{c}{Valid Mock FRBs} \\
         \cline{2-5}
         & Recall & Precision & Recall & Precision \\
    \hline 
         HEIMDALL & 0.884 & \textless0.01 & 0.975 & \textless0.01     \\
         Filter & 0.816 & 0.502 & 0.934 & 0.478     \\
         FETCH & 0.764 & 1.00 & 0.881 & 1.00            \\
    \hline
    \end{tabular}
    \label{tab:recall_precision}
\end{table}

In our setting, HEIMDALL typically yields a few hundreds to a thousand of candidates within the file which contains 81.2 s data. The majority of these spurious candidates are filtered out by the SNR threshold and lower pulse width limit of the candidate filter. The remaining candidates are classified by FETCH, which reached the maximum precision, equal to unity in this test. The overall recall and precision rates for our test are presented in Table \ref{tab:recall_precision}, where we excluded those with an SNR below 10 and a DM exceeding 1500 for valid mock FRBs.

In operation, the typical number of candidates selected after filtering is $\sim10^2$ per day. Instead of the deep learning algorithm \texttt{FETCH}, we can also perform human inspection after filtering at the cost of extra workload and delayed notification for follow-up observations.

\begin{table}[]
    \centering
    \caption{List of bright pulsars in the northern sky, ranked by fluence $F$.   The mean flux density $S$ values are mostly measured at 606 MHz, but those labelled with a $\dag$ are measured at a lower frequency, mostly at 350 MHz. $W_{50}$ is the pulse width at 50\% of peak. $S_{\rm pulse}$ is the mean flux density for each single burst, obtained using $S$ and the duty cycle $W_{50}/{\rm Period}$, $F$ is the corresponding mean fluence of single burst,  $F_{\rm th}$ is the 10$\sigma$ fluence detection limit at the primary beam center of Tianlai, calculated using $W_{50}$ and Equation (\ref{eq:flu_th}).
    Pulsars labeled with * has been detected by the Tianlai Cylinder Pathfinder Array.
    }
    \resizebox{\textwidth}{!}{
    \begin{tabular}{c c c c c c c c}
        \hline
         \multirow{2}{*}{Name} & DM  & \multirow{2}{*}{$S$ (mJy)} & \multirow{2}{*}{Period (s)} & \multirow{2}{*}{$W_{50}$ (ms)} & \multirow{2}{*}{$S_{\rm pulse}$ (Jy)}  & \multirow{2}{*}{$F$ (Jy ms)}  & \multirow{2}{*}{$F_{\rm th}$ (Jy ms)}\\
        & ($\rm pc\,cm^{-3}$) & &&&  &  & \\
        \hline
        B0329+54* & 26.7 & 1337 & 0.714 & 6.6 & 145 & 957 & 100\\
        J0341+5711 & 101 & 364.7$\dag$ & 1.89 & 43 & 16.0 & 689 & 256\\        
        J2238+6021 & 185 & 111$\dag$ & 3.07 & 25 & 13.6 & 341 & 195\\
        B1133+16 & 4.84 & 144  & 1.19 & 5.8 & 29.5  & 171 & 94.1\\
        B2319+60 & 94.6 & 49.5 & 2.26 & 131 & 0.85  & 112  & 430\\        
        B0950+08 & 2.97 & 402  & 0.253 & 8.6 & 11.8 & 102 & 115\\   
        B1919+21 & 12.4 & 72.7 & 1.34 & 34 & 2.87 & 97.6 & 228\\
        J0302+2252 & 19.0 & 70.0$\dag$ & 1.21 & 50 & 1.69 & 84.7 & 276\\
        B2154+40 & 71.1 & 54.7 & 1.53   & 39  & 2.15   & 83.7 & 244\\
        B0809+74 & 5.75 & 51.0 & 1.29   & 41  & 1.60 & 65.8 & 250\\        
        B1237+25 & 9.25 & 46.8 & 1.38   & 45  & 1.44   & 64.6 & 262\\
        B1933+16 & 159  & 161  & 0.359 & 6.3  & 9.17   & 57.8 & 98.1\\        
        B1946+35 & 129  & 78.9 & 0.717  & 19  & 2.98  & 56.6  & 170\\        
        B2021+51* & 22.5 & 95.9 & 0.529 & 7.4 & 6.86  & 50.8 & 106\\
        B1929+10 & 3.18 & 187  & 0.227  & 6.0 & 7.07  & 42.4 & 95.7\\ 
        J2043+7045 & 57.6 & 63.7$\dag$ & 0.588 & 37 & 1.01 & 37.5 & 238\\
        B0823+26* & 19.5 & 62.4 & 0.531 & 5.8 & 5.71   & 33.1 & 94.1\\
        B1859+03 & 402 & 47.4  & 0.655  & 9.0 & 3.45   & 31.0 & 117\\        
        B2020+28* & 24.6 & 59.9 & 0.343 & 12  & 1.71   & 20.5 & 135\\
        B2310+42 & 17.3 & 46.0 & 0.349 & 8.8 & 1.82    & 16.1 & 116\\        
        B0531+21 & 56.7 & 210  & 0.0334 & 3.0 & 2.34   & 7.01 & 67.7\\
        J1022+1001 & 10.3 & 75.0$\dag$ & 0.0165 & 0.97 & 1.28 & 1.24 & 38.5\\
        J0218+4232 & 61.3 & 47.0$\dag$ & 0.00232 & 1.0 & 0.109 & 0.109 & 39.1\\
        \hline
    \end{tabular}
    }
    \label{tab:pulsar}
\end{table}

\subsection{Pulse searching test}
Bright pulsars can be used as target sources to test our pulse searching pipeline.  Table \ref{tab:pulsar} lists bright pulsars, selected from \citet{Pulsar1995} and the ATNF pulsar catalogue \footnote{\text{http://www.atnf.csiro.au/research/pulsar/psrcat}} \citep{ATNF2005}, 
with mean flux density $S\gtrsim 40 \mJy$) at 606 MHz (or 350 MHz if it is not available at 606 MHz)  in the northern sky ($ \delta > 0^\circ$). The objects are listed in a descending order of their mean pulse fluence $F$.
Four of these have been detected by the Tianlai Cylinder Pathfinder Array, which are labeled with a * in Table \ref{tab:pulsar}. Among these, PSR B0329+54 is the only pulsar that is bright enough to be well above our detection limit and can be re-detected readily each day. PSR B2021+51, B0823+26 and B2020+28 have also been detected by our array, probably during their bright phases \citep{Pulsar2023}.  Some of the other pulsars listed in Table \ref{tab:pulsar} may also be detectable.   PSR B1133+16 and B0950+08 have very small DM values, which will be filtered out by our pipeline. The table also give relatively large fluence for PSR J0341+5711 and J2238+6021, but note these are measured at 350 MHz, if scaled to higher frequency their fluence would not be that large.

\begin{figure}
    \centering
    \includegraphics[width=0.99\textwidth]{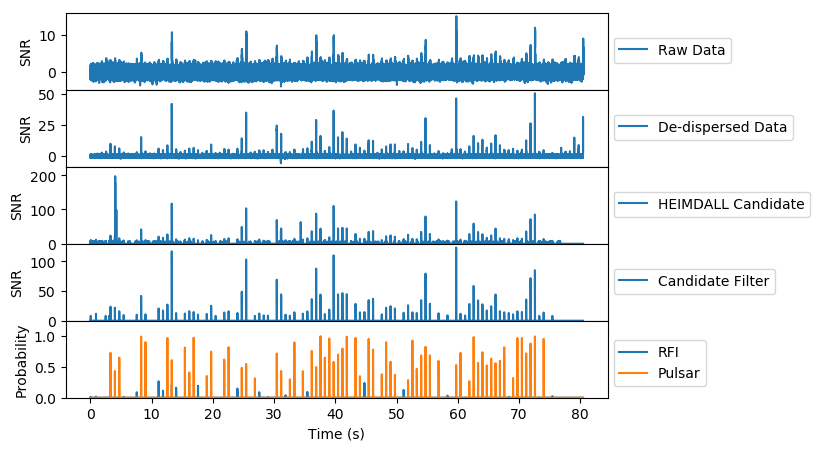}
    \caption{Pipeline tested with PSR B0329+54. Top panel: the raw time series without de-dispersion. The second panel:  the time series data de-dispersed with DM 27.31 $\rm pc\,cm^{-3}$. The third panel: the candidates reported by HEIMDALL. The fourth plot:  the candidates remained after filtering.  Bottom panel: the mean probability of each candidate being astronomical signal provided by the three FETCH models.
    }
    \label{fig:searching_test}
\end{figure}

We use the bright pulsar PSR B0329+54 as a target source to test and evaluate our pulse searching pipeline. Figure \ref{fig:searching_test} shows a segment of the data at the peak of transit of this source. The top panels shows the raw data, the second panel shows the de-dispersed data, where signal is enhanced; the third panel shows the HEIMDALL candidates (note the higher SNR), the fourth panel shows the candidates after filtering, and finally, the bottom panel shows the candidates classified as real signals (in orange) and RFI (in blue).  From this observation data, we derive an average pulse period of 0.7147 s, and a DM value of $27.31 \pc \cm^{-3}$, consistent with those reported in the literature, 0.7145 s and $26.76 \pc \cm^{-3}$ respectively \citep{ATNF2005}.
These results demonstrate that our pipeline successfully identifies most of the pulses from the pulsar as candidates, especially those with higher SNRs. However, it may miss pulses located near strong RFI sources. Also, as noted in Section \ref{sec:de-dispersion and pulse searching}, at the end of each data file there is drop of detection efficiency.  This end-of-file problem can be solved by copying the 8 seconds of data at the end of file to the start of the next data file, so that such missed pulse can still be detected at the next block.

\section{Beam Arrangement and Optimization}
\label{sec-beam-arrange}

In the previous sections we considered detection of a radio pulse with a digitally formed beam. The current digital backend allows the simultaneous formation of a total of 96 beams, which can be pointed to different directions in the sky at will. In this section we study how to arrange these beams, in order to achieve the optimal detection rate for FRBs which occur randomly in the sky, and estimate this rate. We generate a sample of mock FRBs with different fluences and widths, then check whether they can be detected by one of the digitally formed beams. The arrangement of the beams are varied to find the optimal configuration. 

\subsection{FRB distribution}
\begin{figure}
    \centering
    \includegraphics[width=0.33\textwidth]{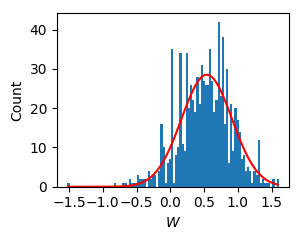}
    \includegraphics[width=0.33\textwidth]{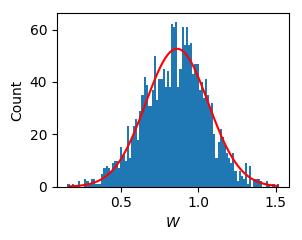}
    \includegraphics[width=0.33\textwidth]{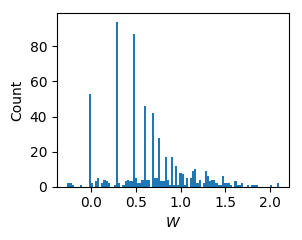}
    \caption{Left: Histogram of the adopted FRB width in Blinkverse FRB catalog with normal distribution fitting shown in red line. Middle: Histogram of the FAST sample with normal distribution fitting shown in red line. The FAST sample include mainly thousands of bursts from only a few active repeaters, which may not be representative for the overall distribution. FAST sample have larger $\mu_{w} = 0.8626$ and smaller $\sigma_{w} = 0.2020$. Right: Histogram of excluded CHIME samples. CHIME samples are excluded for the majority of them are given in only several values.}
    \label{fig:width_distribution}
\end{figure}

To generate a sample of FRBs, we assume its cumulative fluence distribution follows\citep{CHIME202112}
\begin{equation}
D_{\rm Flu} (> F) =  R_{\rm sky, 5\,Jy\,ms} \left(\frac{F}{5 \Jy\ms}\right)^{\alpha}
\label{eq:str_distribution_cu}    
\end{equation}
The all-sky FRB rate $R_{\rm Sky, 5 \Jy \ms }= 525$ per day at the relevant frequency band, and $\alpha=-1.4$.
We grid the fluence distribution from 0.25 to 2000.25 $\Jy\,\ms$ evenly into 4000 intervals, with $\Delta F = 0.5 \Jy\,\ms$. The FRB all sky rate within a interval is then given by
\begin{eqnarray}\label{eq:str_distribution_diff}
\Delta D_{\rm Flu}(F)=D_{\rm Flu} (F) - D_{\rm Flu} (F+\Delta F) .
\end{eqnarray}
For the pulse width We fit a log-Normal distribution for the sample from the Blinkverse FRB catalogue\citep{Blinkverse2023}, as Figure   \ref{fig:width_distribution} (left) shows. In that catalogue, we exclude the CHIME and FAST FRBs, as the CHIME FRBs are given just in a few values (Figure \ref{fig:width_distribution} right), while the FAST FRBs are thousands of bursts from only a few active repeaters, which may not be very representative for our sample (Figure \ref{fig:width_distribution} middle). From Figure \ref{fig:width_distribution}, we can see that the FAST and CHIME samples also follow a normal distribution with similar $\mu_{w}$, thus excluding them will not cause major impact on our results. The probability density function of the fitted distribution is given by:
\begin{equation}
D_{w} (W) = \frac{1}{\sqrt{2\pi} \sigma_{w}}\exp(-\frac{(W-\mu_{w})^2}{2\sigma_{w}^2})
\end{equation}
where $W = \log_{\rm 10}(\Delta t/\ms)$,  $\mu_{w} = 0.5386$ and $\sigma_{w} = 0.3732$. We then grid $W$ from --1.5 to 2.5 into 1000 intervals, with a $\Delta W = 0.004$.

\begin{figure}
    \centering
    \includegraphics[width=0.6\textwidth]{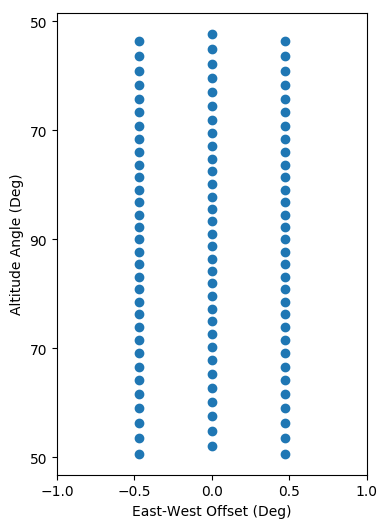}
    \caption{The adopted beam arrangement, each point marks the center of a digital beam. The plot here shows the optimal configuration, with a spacing  of $2.27\degree$ in north-south and $0.47\degree$ in east-west respectively.}
    \label{fig:beam_center}
\end{figure}

\subsection{Beam Arrangement}
Although a digitally formed beam can be pointed at any direction in the sky, obviously it should be placed within the main lobe of the primary beam to be effective. As the primary beam of our cylinder antenna is a long strip in the north-south direction, we place the digital beams in three parallel rows across the East-West direction, and on each row 32 beams are lined along the North-South direction, with constant E-W and N-S spacings, as shown in Figure \ref{fig:beam_center}.

Following \citet{2015PhRvD..91h3514S} and \citet{Tianlai2023,Yu:2023idm}, we model the voltage response as a product of the North-South distribution and the East-West distribution, 
\begin{equation}
A(\hat{\vb*{n}}) = A_{\text{NS}} (\sin^{-1}(\hat{\vb*{n}} \cdot \hat{\vb*{x}}); \theta_{\text{NS}}) \times A_{\text{EW}} (\sin^{-1}(\hat{\vb*{n}} \cdot \hat{\vb*{y}}); \theta_{\text{EW}}) 
\end{equation}
where $\hat{\vb*{x}}$ and $\hat{\vb*{y}}$ are the unit vector pointing to East and North, respectively, and the NS and EW functions are 
\begin{equation}
A_{\text{NS}} = \exp \left[ - 4 \ln2 \left( \frac{\theta}{\theta_{\text{NS}}} \right)^2 \right],
\label{eq:beam-ns}
\end{equation}
where $\theta_{\text{NS}}(\nu) = \alpha_{\text{NS}} \frac{\lambda}{D_{\text{NS}}}$, with $D_{\text{NS}} = 0.3$m;
\begin{equation}
A_{\text{EW}} \propto 
     \int^{\frac{W}{2}}_{-\frac{W}{2}}A_D\left(2 \tan^{-1}\left(\frac{x}{2FW}\right) ;\theta_\text{EW}\right) 
      e^{-i\frac{2\pi}{\lambda}x\sin\theta} \dd x, 
    \label{eq:beam-ew}
\end{equation}
where 
\begin{equation}
    A_{D}(\theta; \theta_{\text{FWHM}}) = \exp \left[ -\frac{\ln 2}{2} \frac{\tan^2\theta}{\tan^2(\theta_{\text{FWHM}}/2)} \right],
\end{equation}
$W=15$ m, and fitting the parameters $\alpha_{\text{NS}}$, $F_R$ and $\theta_{\text{EW}}$ to the results of electromagnetic field simulation \citep{Sun_2022}, yields $\alpha_{\text{NS}} = 1.04$, $F_R = 0.2$ and $\theta_{\text{EW}} = 2.74$.

\begin{figure}
    \centering
    \includegraphics[width=0.45\textwidth]{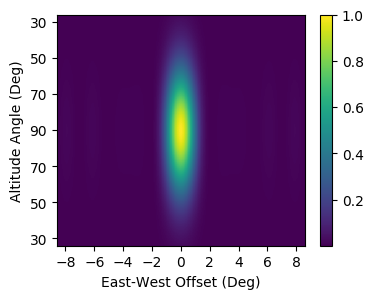}
    \includegraphics[width=0.485\textwidth]{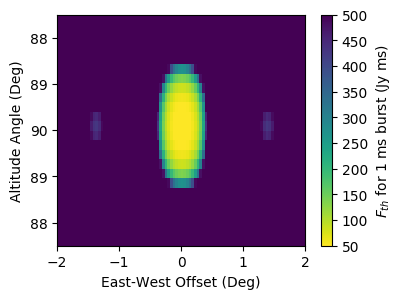}
    \caption{Left: The adopted primary beam profile for the cylinder array. Right: The profile of an effective digital beam.}
    \label{fig:prim_profile}
\end{figure}

The effective digital beam will be a product of the primary beam and the array form factor. For the cylinder array, both the primary beam and the digitally formed beam are elongated along the N-S direction, as shown in Figure \ref{fig:prim_profile}. 

Each digital beam has a $\sim 0.5^\circ$ width in the E-W direction, so the three rows of digital beam could cover about $1.5^\circ$ if they do not overlap each other, while the  primary beam has a width of $\sim 2^\circ$. As shown in Figure \ref{fig:prim_profile}, the digital beams have multiple side lobes along the East-West direction. This is unavoidable as the baselines between feeds on different cylinders are longer than the half-wavelength. These lobes are however modulated by the primary beam, so that they are attenuated as we move off the center of primary beam. The presence of these lobes could cause some ambiguity in the localization of a burst, but this ambiguity could be partly broken by jointly using the signal strength of multiple beams. 
To reduce the gap between beams, we shift the beams in the center row by a half of the N-S spacing towards South, so that the tip of beams in the two adjacent rows fit nicely into the spacing between the beams in the central row, forming a closely packed structure.

The detection rate function is constructed as follows. First we grid the sky evenly in a topocentric Cartesian coordinate system. The North-South component of the unit vector, from --0.9 to 0.9, is gridded into 501 pixels, corresponding to altitude angle greater than $25.8^\circ$. The East-West component of the unit vector, from --0.15 to 0.15, is gridded into 401 pixels, corresponding to the E-W offset within $\pm 8.6^\circ$. We obtain the beam response $B_{ij,k}$ of the $k$th beam at pixel $(i,j)$ by multiplying the synthesized beam profile centered at its given direction with the primary beam profile. The fluence threshold is estimated as
\begin{equation}
{F_{\rm th}} = N_{\rm th}S_{\rm n}\Delta t = N_{\rm th} S_{\rm cal} \sqrt{\tau \Delta t} \frac{B_{\rm cal}}{B_n} \frac{\sigma_T}{\Delta_{\rm cal}} 
\label{eq:sensitivity}    
\end{equation}

\begin{figure}
    \centering
    \includegraphics[width=0.49\textwidth]{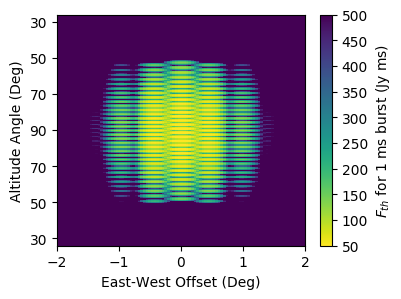}
    \includegraphics[width=0.49\textwidth]{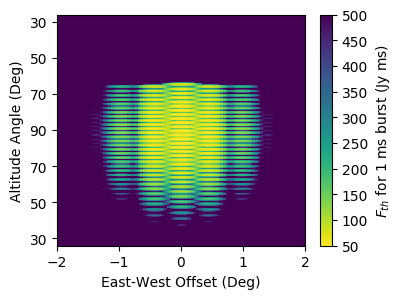}
    \caption{The Overall beam profile $R_{ij}$, given in terms of the detection threshold $F_{\rm th}$.
    Left: the case where the center is at the Zenith; Right: the case where the center is shifted $10^\circ$ to north have more coverage on the North Celestial Pole. 
    }
    \label{fig:beam_profile.png}
\end{figure}

\begin{figure}
    \centering
    \includegraphics[width=0.34\textwidth]{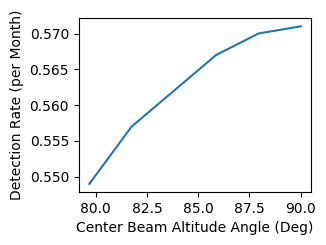}
    \includegraphics[width=0.32\textwidth]{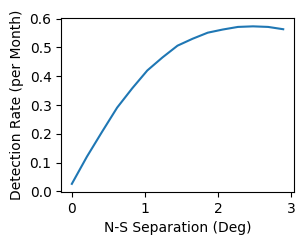}
    \includegraphics[width=0.33\textwidth]{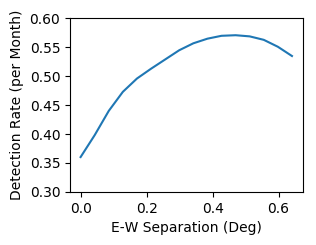}
    \caption{detection rate with respect to the center beam's altitude angle (left) and beam separation in N-S (middle) and E-W (right) direction. The optimal values are $90\degree$, $2.27\degree$ and $0.47\degree$, respectively.} 
    \label{fig:detection_depending}
\end{figure}

\begin{figure}
    \centering
    \includegraphics[width=0.6\textwidth]{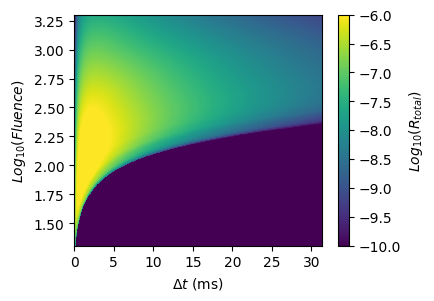}
    \caption{Estimation of the fluence distribution of observed FRBs. 
    }
    \label{fig:fluence_distribution.png}
\end{figure}

For a given point in the sky, one of the digitally formed beams will have the most sensitive response. This digital beam can be identified by finding the maximum of $B_k(\vec{n})$, though on some exceptional points (boundary) more than one beam may have equal sensitivity. In more sophisticated algorithms, the data from more than one beam may be combined to enhance the sensitivity, but this will increase the complexity of the algorithm, and in most cases the actual enhancement in sensitivity is slight, though the localization of the burst can be improved by the joint fit of multiple beams. Below for the sensitivity estimate we will use the one best digital beam and its corresponding threshold, with SNR threshold $N_{\rm th}$ set to 10 to estimate our detection rate. We then have 
\begin{equation}
F_{\rm th}= 12.58 ~N_{\rm th} \sqrt{\tau \Delta t}  \left(\frac{B_{\rm cal}}{\max\limits_{k}(B_{ij,k})}\right)~ \Jy \ms
\label{eq:flu_th}
\end{equation}
The FRB detection rate of a pixel is then estimated by
\begin{equation}
{R_{ij}} =  \frac{A_{ij}}{(360^2/\pi)}\sum\limits_{F,\Delta t}\limits^{F>F_{th}(\Delta t)}{\Delta D_{\rm Flu}(F)D_w (\Delta t)}
\label{eq:frb_rate}    
\end{equation}
where i,j denotes the pixel coordinate, $A_{ij}$ is the size of pixel in the unit of squared degrees, and the whole sky area is $360^2/\pi~{\rm deg}^2$.
$R_{ij}$ is plotted in Figure \ref{fig:beam_profile.png}. The overall FRB detection rate is obtained by summing over the contribution from all pixels. 

In Figure \ref{fig:beam_profile.png}, the left panels shows an arrangement with the center of the digital beams at the zenith, which best utilized the most sensitive part of the primary beam. In the right panel, we showed any arrangement where the center is shifted $10^\circ$ northward, so that the north celestial pole (NCP) which has an altitude angle of $44^\circ$ is better covered. However, due to the limited number of beams, the southern part of the sky is then less well covered, and because for this cylinder array the sky area with lower altitude generally have lower sensitivities, the overall detection rate will be reduced. 

To optimize our FRB detection rate, we adjust the pointing the digitally formed beams, trying different separation between the rows of beams, and the separation between beam centers between each column. The optimization is achieved by maximizing a function that takes these parameters as input and returns the estimated FRB detection rate.
We search for the maximum of the detection rate function by varying the beam separation in the range of $(0, 14)$ and $(0,15)$ pixels for the beam separation between and within columns, respectively, with a starting value of 2 for both parameters. 

In Figure \ref{fig:detection_depending}, we show how the detection rate varies with altitude of the center beam, the N-S beam separation, and the E-W direction beam separation. These parameters are varied one at a time, with the other two parameters fixed to the optimal value. 
As we can see from the left panel of Figure \ref{fig:detection_depending}, the highest detection rate is achieved when the center of array of the digitally formed beams is aimed at the zenith, which is the peak of the primary beam of the cylinder, thus having the highest sensitivity. If we shift the center away from the zenith, the total detection rate would be lowered.

Varying the separation of the digitally formed beams in either the N-S or E-W direction have two effects: on the one hand, increasing the beam separation reduces the beam overlap, and allow the digitally formed beams to cover a larger total sky area, which increases the rate of FRB detection; on the other hand, this also means that some of the digitally formed beams are placed off-center, leaving holes in the main lobe of the primary beam where the sensitivity is the highest. The optimal detection rate is achieved at a point where these two trends are balanced, and most of the crest of the primary beam is covered. We find that for the optimal arrangement, the beam separation is $0.47\degree$ in the E-W direction and $2.27\degree$ in the N-S direction near the center of FoV. At this optimal configuration, the detection rate is 0.571 FRBs per month for a $10\sigma$ burst.

In Figure \ref{fig:fluence_distribution.png}, we show the distribution of the fluence and width of the detected bursts. There is a clear lower limit in the fluence of the detected bursts, which increases with increasing width of the burst, which we already noted in the discussion of the sensitivity in Section \ref{sec:sensitivity}. A high concentration of the detected bursts will have a short burst width of a few ms, with fluence less than 100 Jy ms, but of course there will also be many detectable bursts.   
This result is somewhat dependent on the assumed model of FRB fluence and duration distribution, but it shows what kind of FRBs we expect to detect.

\section{An FRB detection}
\begin{figure}
\centering
\includegraphics[width=0.6\textwidth]{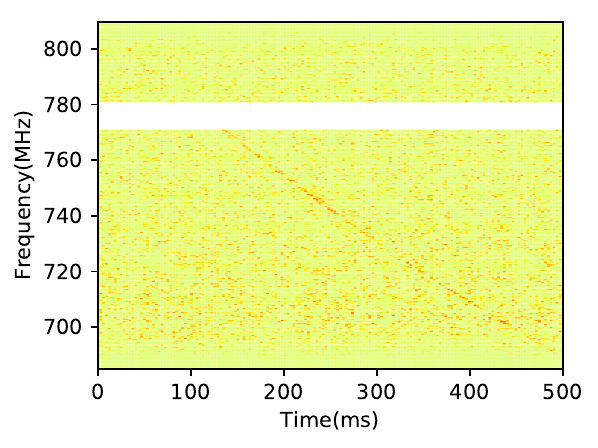}
\caption{The raw waterfall plot of beam 41 for FRB 20220414A. 
}
\label{fig:FRB20220414waterfall}
\end{figure}

\begin{figure*}
\centering
\includegraphics[width=0.33\textwidth]{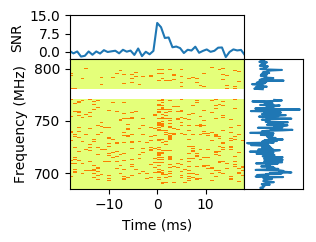} 
\includegraphics[width=0.33\textwidth]{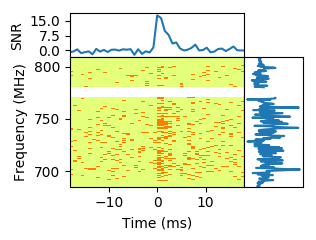}
\includegraphics[width=0.33\textwidth]{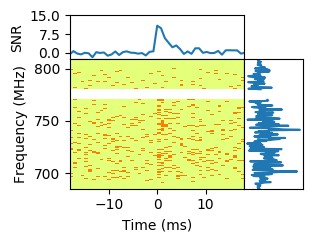}\\
\includegraphics[width=0.33\textwidth]{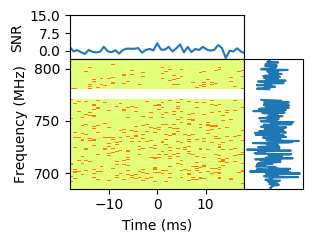}
\includegraphics[width=0.33\textwidth]{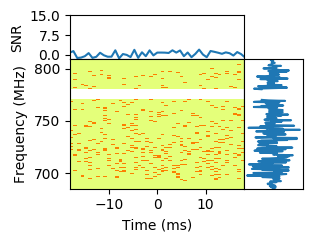}
\caption{The de-dispersion plot of FRB 20220414A. Top: the three beams 40 (top left) 41 (top middle), 42 (top right); Bottom: the two neighboring beams where it is not detect: 39 (bottom left), 43 (bottom right). }
\label{fig:FRB20220414}    
\end{figure*}

During test observation, We detected a bright FRB, designated as FRB 20220414A, at UT 17:26:40.368 on April 14, 2022. In this observation, we did not use the optimal configuration of beam arrangement discussed above, instead the 96 beams were arranged in a single row along the meridian, with Alt $\geq 60^\circ$. The separation between beams was $\approx 0.6316^\circ$. The FRB was detected in three neighboring beams (\#40, \#41, \#42) with SNR = 11.9, 17.7 and 10.9, respectively. In Figure \ref{fig:FRB20220414waterfall}, we show the raw frequency-time waterfall plot of the beam 41. The band at 770--780 MHz is contaminated by an RFI (probably from the newly allocated 5G mobile signal) and is removed. In Figure \ref{fig:FRB20220414} we plot the de-dispersion result for the three beams where it is detected, as well as two neighboring beams where it is undetected.  The detection was reported on the Astronomer's Telegram \citep{Yu2022}, and the Transient Name Server (TNS) \footnote{\url{ https://www.wis-tns.org/object/20220414a}}. 

For a burst in the direction $\bm{q}$ with fluence $F$, the observed data of synthesized beam $\alpha$ is 
$$F_\alpha = F B_\alpha(\bm{q})+\eta_{\alpha},$$ where $\eta_{\alpha}$ is the noise, $F B_\alpha(\bm{q})$ is the expected signal, and the response of the synthesized beam $B_\alpha(\bm{q})$ is given by
\begin{equation}\label{eq:beam_loc}
    B_\alpha(\bm{q}) = \left|A(\bm{q})\sum_a \exp(-2\pi j(\bm{q}-\bm{k}_\alpha)\cdot \bm{u}_a)\right|^2,
\end{equation}
where $\bm{k}_\alpha$ is the unit vector for the direction of the synthesized beam $\alpha$, and $A(\bm{q})$ is the voltage response of the primary beam.


The burst can be located by minimizing the likelihood.
We adopt a reduced log likelihood,
\begin{eqnarray}
    \ln\mathcal{L}(\bm{q}, F_s) &=& -\sum_\alpha \frac{(F_s B_\alpha(\bm{q})-F_\alpha)^2}{2\sigma_T^2\Delta t_s/\tau}\nonumber\\
    &=& -\frac{1}{2} \sum_\alpha \left(\frac{F_s B_\alpha(\bm{q})}{\sigma_T\sqrt{\Delta t_s/\tau}}-{\rm SNR}_\alpha\right)^2
\end{eqnarray}
where $F_{\alpha}$ is the flux detected by the beam $\alpha$, $F_s$ is the model fluence of the FRB, and $\Delta t_s$ is the observed burst width. The second line is written in terms of the SNR, ${\rm SNR}_\alpha$ for beam $\alpha$. Only beams with significant SNR would contribute significantly to this likelihood. 
In principle, equation (\ref{eq:beam_loc}) is frequency dependent. However, the main-lobe does not vary significantly with frequency, therefore we average $B_\alpha(\bm{q})$ over 700--800 MHz and use the averaged synthesized beam for likelihood.

As mentioned in Section \ref{sec-beam-arrange}, there are multiple sidelobes in the East-West direction. In the present case, the beams are aligned only in the North-South direction, so the location is even more uncertain in the EW direction. However, due to the primary beam of the antenna in the EW direction, if the FRB occurs in a sidelobe, it would be very bright in order to be detected, and in that case it might also be seen in other beams. The fact that it is not detected by the other beams itself place some constraint on its location in the EW. 

\begin{figure}
\centering
\includegraphics[width=0.6\textwidth]{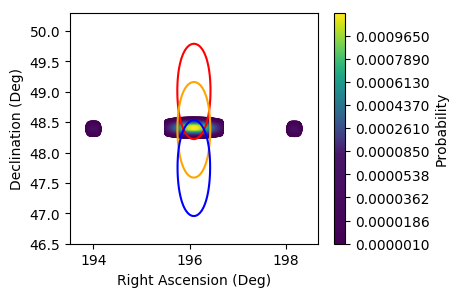}
    \caption{The FRB 20220414A localisation probability. The red, orange and blue circles labels the --3 dB gain range of beam 40, 41 and 42, respectively, the region with probability less than the minimum of the color bar is shown as white. 
    }
    \label{fig:FRB220414A_loc}
\end{figure}

\begin{figure}
    \centering    \includegraphics[width=0.6\textwidth]{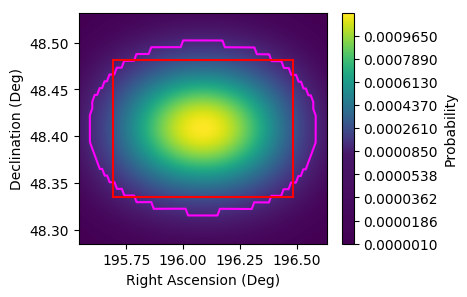}
    \caption{The main lobe of Figure \ref{fig:FRB220414A_loc}. The magenta ellipsoid indicates the 90\% cumulative probability region. The red rectangle indicates the 90\% cumulative probability region obtained from the 1-d cumulative probability of RA and Dec, respectively. 
    }
    \label{fig:loc_zoom}
\end{figure}

The localisation may be enhanced if we adopt a prior on the fluence $F_s$ using the fluence distribution in equation (\ref{eq:str_distribution_diff}),
\begin{equation}
\ln L_{\rm Flu}(F_s) = \ln\Delta D_{\rm Flu}(F_s)
\end{equation}
which helps to improve our localization in the EW direction.
The probability of the location of the burst is shown in Figure \ref{fig:FRB220414A_loc}, where the red, orange and blue circles show the --3 dB range from the peak of the \#40, \#41, and \#42 beams. The pixels with probability less than the minimum of the color bar, i.e.  $10^{-6}$, are shown in white. We can see that the most likely position are at the overlapping regions of the three beams, though there are also sidelobe areas to the east and west of the main lobe region. A zoom up plot of the main lobe is shown in Figure \ref{fig:loc_zoom}, where the ellipse show the region of 90\% probability, while the rectangle marks the 90\% cumulative probability region obtained from the 1-d RA and Dec probability. With the information from multiple beam detection, the burst is localized with a precision better than the synthesized beam size. 

The properties of this FRB are listed in Table \ref{tab:20220414_properties}. The best fit total DM is $207.6 \pc \cm^{-3}$, slightly lower than we originally reported in Astronomer's Telegram. We can infer the redshift of the burst using the DM-redshift relation \citep{Zhang2018}, 
\begin{equation}
z_{max} \sim \frac{\rm DM_E}{855\,\rm pc\,cm^{-3}}\,(z_{\rm max}<3).
\end{equation}
The total DM is given by
\begin{equation}
    \rm DM_{\rm total} = DM_{\rm MW} + DM_{\rm halo} + DM_{\rm IGM} + \frac{DM_{\rm host}}{1+z}
\end{equation}
where $\rm DM_{\rm MW}$ is the Galactic contribution, which can be estimated using the NE2001 and YMW16 models \citep{Cordes2002,Yao2017,Price2021}. 

Assuming $\rm DM_{\rm MW} = 21.1\,\rm pc\,cm^{-3}$, and neglecting the unknown contribution from the halo and the host, we derive an upper limit for the DM-inferred redshift $z_{\rm max}\leq 0.24$. On the other hand, if we adopt a galactic halo contribution \citep{Dolag2015,Prochaska2019}, 
$\rm DM_{\rm halo} \sim 50\,\rm pc\,cm^{-3},$  
the extragalactic DM would then be estimated as
\begin{equation}
{\rm DM_{E} = DM_{\rm IGM} + \frac{DM_{\rm host}}{1+z} = 136.5\,\rm pc\,cm^{-3}},
\end{equation}
leading to an upper value for the redshift $z_{\rm max}=0.16$. Adopting the \citep{Planck2016} model, the distance is then 
$$D_L \leq \frac{c z_{\rm max}}{H_0} = 708\,\rm Mpc,$$ luminosity 
$$L_p \leq 4\pi D_L^2 S_{\rm peak} \nu_c = 5.76 \times 10^{43}\, \rm erg\,s^{-1}.$$

Despite the large fluence and peak flux recorded, its luminosity is estimated to be at an average level for one-off FRBs due to its relatively small DM. A 2-hour follow-up observation was conducted by FAST on April 22, but no additional burst was detected.

\begin{table}
    \centering
    \caption{Properties of FRB 20220414A.}
    \begin{tabular}{c c}
    \hline 
         RA  &  13h 04m 21s ($\pm$ 1m35s)\\
         DEC &  $+48^\circ 24' 32''$ ($\pm 4'25''$)\\
         DM & $207.6 \,\rm pc\,cm^{-3}$\\
         Galactic DM (YMW16) & $21.1\,\rm pc\,cm^{-3}$\\
         Galactic DM (NE2001) & $27.5\,\rm pc\,cm^{-3}$\\
         DM-inferred Redshift & $\leq0.24$\\
         Peak flux density & $128.4\, \Jy$ \\
         Duration & $2.2\, \ms$\\
         Fluence & $204.1\, \Jy\, \ms$ \\
         Estimated redshift & $< 0.16$\\
         Luminosity & $\leq 5.76 \times 10^{43}\, {\rm erg\,s}^{-1}$ \\
    \hline
    \end{tabular}
    \label{tab:20220414_properties}
\end{table}

We also searched for possible host galaxy in the SDSS DR17 spectroscopic catalog \footnote{https://skyserver.sdss.org/dr17/SearchTools/SQS}. The RA and Dec is constrained to be within the 90\% confidence interval given in Table \ref{tab:20220414_properties} and the redshift is lower than 0.24. We found 8 optic galaxies and the information is listed in Table \ref{tab:sdss_opt_host}.

\begin{table*}
    \centering
    \caption{Optic galaxies in SDSS DR17 spectroscopic catalog with RA and Dec inside the 90\% confidence interval and redshift lower than 0.24.}
    \begin{tabular}{c c c c}
         Object & RA & Dec  & Redshift \\
    \hline 
         SDSS J130556.97+482013.6 & 13h 05m 56.97s & $+48^\circ 20'13.6''$ & 0.141\\
         SDSS J130524.90+482110.2 & 13h 05m 24.90s & $+48^\circ 21'10.2''$ & 0.181\\
         SDSS J130535.68+482000.9 & 13h 05m 35.68s & $+48^\circ 20'00.9''$ & 0.154\\
         SDSS J130417.23+482651.7 & 13h 04m 17.23s & $+48^\circ 26'51.7''$ & 0.182\\
         SDSS J130449.78+482740.1 & 13h 04m 49.78s & $+48^\circ 27'40.1''$ & 0.181\\
         SDSS J130327.90+482302.0 & 13h 03m 27.90s & $+48^\circ 23'02.0''$ & 0.135\\
         SDSS J130342.72+482207.2 & 13h 03m 42.72s & $+48^\circ 22'07.2''$ & 0.152\\
         SDSS J130321.31+481950.9 & 13h 03m 21.31s & $+48^\circ 19'50.9''$ & 0.008\\
    \hline
    \end{tabular}
    \label{tab:sdss_opt_host}
\end{table*}

\section{Conclusion}

This paper presents the design and implementation of the FRB-search  backend for the Tianlai cylinder pathfinder array, as well as its performance obtained from simulations and test observation campaign. 
We described the beam forming principle, the hardware setup, as well as the pipeline for incoherent de-dispersion and candidate pulse search and classification. The present backend is capable of generating a total of 96 digital beams, covering approximately 40 square degrees of the sky (3 dB below the peak). This is smaller than the area of the primary beam, larger area could be covered if the computing power is increased to process a larger number of digitally formed beams. 
We have estimated the fluence detection threshold which varies with the pulse width, and depends also on its position.  The performance of the backend has been estimated through mock FRBs injection and validated through observation of pulsars. 

We have optimised the digital beam placement by maximising the FRB detection rate, which has been estimated to be around $\sim 0.5$ FRB event per month with the current 96 beam setup. However, this rate estimation and the optimized configuration are subject to the primary beam shape and FRB parameter distribution, both of which are poorly known at present.Nevertheless, it provide some approximated estimate that is useful for planning our experiment.

After conducting a four months FRB search campaign we realised our first discovery, namely FRB 20220414A, which we were able to localise with a precision of $\sim 0.1^\circ$. 
We are currently working on adding new outrigger cylinders, which would improve significantly the array localisation capability. 

\begin{acknowledgements}
The Tianlai Cylinder Pathfinder Array Transient Digital Backend is built with the funding from the Repair and Procurement Grant of the Chinese Academy of Science. For this research we acknowledge the support of the National SKA program of China ( Nos. 2022SKA0110100 and 2022SKA0110101), the National Natural Science Foundation of China (Nos. 1236114814, 12203061, 12273070, 12303004).
\end{acknowledgements}

\bibliographystyle{raa}
\bibliography{ms2024-0097refs}

\end{document}